\documentclass{iaus}
\usepackage{graphicx}
\usepackage{amssymb}
\usepackage{natbib}

\newcommand{\msun}{$\mbox{M}_{\odot}$}

\title[Star-Forming Rings in Spiral Galaxies] 
{A SINFONI view of circum-nuclear star-forming rings in spiral galaxies}

\author[Falc\'on-Barroso et al.]   
{Jes\'us Falc\'on-Barroso$^{1}$, 
Torsten B\"oker$^{1}$,
Eva Schinnerer$^{2}$,\\
Johan~H. Knapen$^3$,
Stuart Ryder$^4$}

\affiliation{$^1$European Space and Technology Centre, Keplerlaan 1, 
2200 AG Noordwijk, The Netherlands \break email: jfalcon@rssd.esa.int\\[\affilskip]
$^2$Max-Planck Institut f\"ur Astronomie, K\"onigstuhl 17, D-69117 Heidelberg, Germany\break
$^3$Instituto de Astrof\'isica de Canarias, Via L\'actea s/n, E-38200 La Laguna, Spain\break
$^4$Anglo-Australian Observatory, PO Box 296, Epping, NSW1710, Australia}

\pubyear{2007}
\volume{245}  
\pagerange{X--X}
\date{?? and in revised form ??}
\jname{Proceedings Title IAU Symposium}
\editors{M. Bureau, E. Athanassoula, \& B. Barbuy, eds.}
\begin{document}

\maketitle

\begin{abstract}
We present near-infrared (H- and K-band) {\tt SINFONI} integral-field 
observations of the circumnuclear star formation rings in five
nearby spiral galaxies. We made use of the relative intensities of different
emission lines (i.e. [Fe{\sc II}], He{\sc I}, Br$\gamma$) to age date the stellar 
clusters present along the rings. This qualitative, yet robust, method allows us 
to discriminate between two distinct scenarios that describe how star formation 
progresses along the rings. Our findings favour a model where star formation 
is triggered  predominantly at the intersection between the bar major axis and 
the inner Lindblad resonance and then passively evolves as the clusters 
rotate around the ring ({\it 'Pearls on a string'} scenario), although models of 
stochastically distributed star formation ({\it 'Popcorn'} model) cannot be 
completely ruled out.\looseness-2

\keywords{galaxies: spiral; galaxies: nuclei; galaxies: stellar content; 
          galaxies: star clusters; galaxies: starburst; galaxies: individual 
	  (NGC\,613, NGC\,1079, NGC\,1300, NGC\,5248, IC\,1438)}
\end{abstract}

\firstsection 
\section{Introduction}

Star-forming nuclear rings in the inner regions of early- and intermediate 
Hubble type (Sa-Sc) galaxies are one of the key pieces to understand 
the transport of gas to the nuclei of galaxies and therefore how secular 
evolution proceeds. The basic picture of how these rings form seems well 
established, both theoretically and observationally, and is usually associated to 
the interplay between bar-driven inflow and bar resonances (e.g. 
\citealt{schwarz84, cg85, atha92, knapen95}). A non-axisymmetric 
gravitational potential, nearly always due to the presence of a stellar bar or oval 
distortion, causes the gas to lose angular momentum and spiral in towards the 
nucleus. During the inflow, and because of its dissipative nature, the gas 
accumulates around the radii at which the stellar orbits experience dynamical 
resonances with the rotating bar potential. In the case of the nuclear rings 
discussed here this typically happens at the Inner Lindblad resonance. When 
observed in more detail, the gas enters the ring via two tightly wound spiral 
arms or dust lanes. At the contact points between the dust lanes and the ring, 
the gas becomes less turbulent, and enters the almost circular orbits delineating 
the ring.  While it is clear that there is abundant (molecular) gas throughout the ring, 
there is some debate about how and where star formation occurs. 

In this paper we present our efforts to understand the preferred model of 
star formation along the rings, based on {\tt SINFONI} integral-field observations, 
of circumnuclear star-forming rings in five nearby spiral galaxies (NGC\,613, 
NGC\,1079, NGC\,1300, NGC\,5248, IC\,1438). We refer the reader to 
\cite{boker07} for a more complete discussion of the full sample.

\begin{figure}[t]
\begin{center}
\includegraphics[angle=0, width=.99\textwidth]{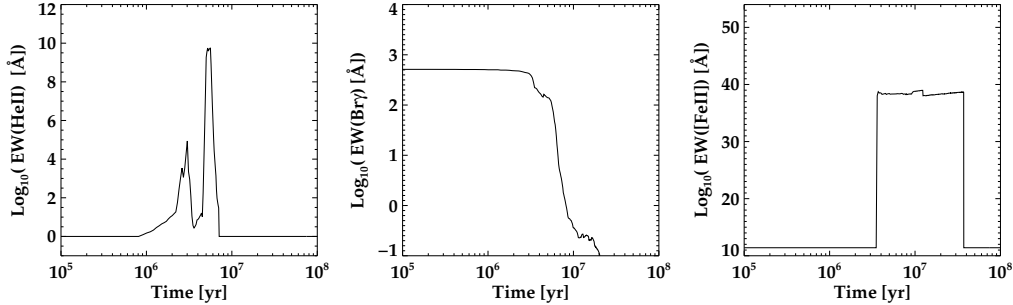}
\caption{Evolution of the equivalent widths of  He{\sc II}, Br$\gamma$ and 
[Fe{\sc II}] as a function of time for instantaneous STARBURST99 models. The
maximum contribution of each of the three lines occurs in distinct, yet
overlapping, time episodes. We use this property to produce a more accurate 
age dating of the stellar clusters along the star-forming rings (see text). Data
shown here are  for models with z=0.02, an IMF slope $\alpha$=2.35,
with an upper mass limit of 100 \msun. Models for other metallicities, IMF
slopes and upper mass limit show similar behaviour.\looseness-2}
\end{center}
\end{figure}

\section{Measuring the stellar cluster ages}
Measuring the relative ages of the different stellar clusters (i.e. hotspots) along 
the ring is a difficult task. The orbital time scales in the ring are short (e.g. a few 
tens of Myr). In order to perform the age dating it is therefore imperative that 
one uses a tracer that it's sensitive to these time scales. The most widely used 
diagnostic for this purpose is the H$\alpha$ emission line. There are several 
reasons why this line is a popular choice. First, it is an optical emission line that 
can be easily measured. Second, its equivalent width (EW) decreases almost 
monotonically with time for an instantaneous burst (see the STARBURST99 
prediction, \citealt{sb99}) allowing us determine the evolution stage of a 
cluster depending of the EW value. The use of this emission line, however, is 
limited to clusters in the age range 3 to 10 Myr, and therefore any inferred age 
differences have to be small (few Myrs). Below 3 Myr the EW of the H$\alpha$ is 
almost constant, and above 10 Myr the value is well below typical detection 
limits ($\sim$1\AA).

An alternative approach is to make use of several spectral lines, whose emission 
peak at different times. In our case we use the flux of three emission lines that 
are prominent in the NIR spectra of the star forming regions in our sample of 
galaxies: He{\sc I}, Br$\gamma$, and [Fe{\sc II}]. Under the assumption that 
the underlying bulge/disk can be considered quiescent, and that most of the 
emission is produced in the ring itself, line fluxes are a better probe than the 
EWs because they require an accurate knowledge of the continuum emission not only 
from the hot spot itself but also the underlying bulge and/or disk, which is very 
hard to measure. The He{\sc I} and Br$\gamma$ lines are both produced by 
photo-ionization in the vicinity of O- or B-type stars. Given that the ionization 
energy for the He{\sc I} line is higher than that of Br$\gamma$, it requires the 
presence of hotter and more massive stars, and hence its flux falls off more 
rapidly after an instantaneous burst than that of the Br$\gamma$ line. The time 
range covered by these two lines is from 0 to 10 Myr. Larger ages can be probed 
with the  [Fe{\sc II}] line (a tracer of fast shocks produced in supernova remnants) 
whose contribution is almost constant from 3 to 35 Myr, where it sharply 
decreases. With these three lines we are therefore able to probe ages
in the range between 0 and 35 Myr, which represents a good match to the 
expected travel time of gas and star clusters around the ring. An illustration of 
their evolution with time (as predicted by  STARBURST99) is presented in Figure~1.
Note that in the figure He{\sc II} ($\lambda \lambda$4686\AA) has been used 
as a surrogate for He{\sc I}, and [Fe{\sc II}] ($\lambda \lambda$1.26$\mu$m)
for [Fe{\sc II}] ($\lambda \lambda$1.64$\mu$m).

\begin{figure}[t]
\begin{center}
\includegraphics[angle=270, width=.99\textwidth]{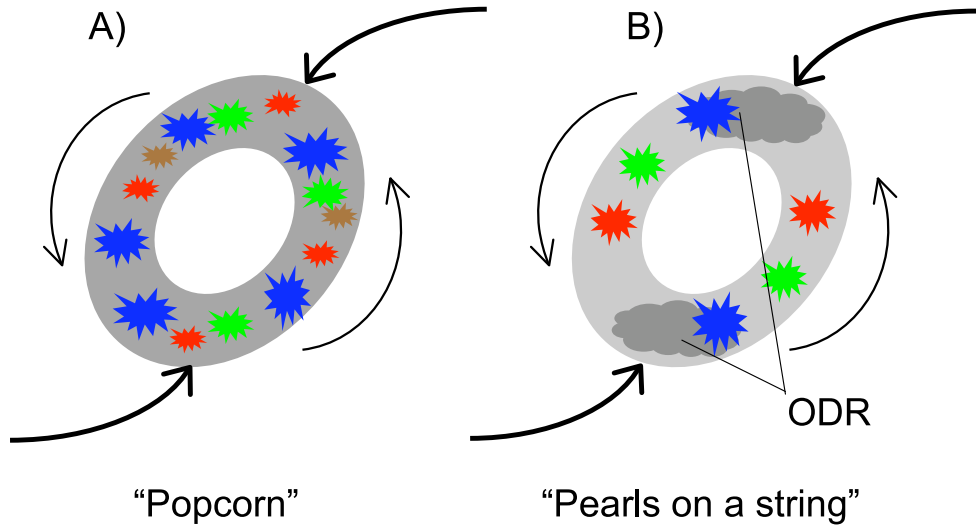}
\caption{Illustration of two proposed scenarios for the evolution of star 
formation in a nuclear ring. Dark grey areas denote dense, cold gas that is 
likely to lead to star formation. The various star symbols denote young stellar
clusters, their colors reflect an increase in cluster age from blue to green to
red. A clear age sequence is  expected only in the {\it 'Pearls on a string'}
scenario. Figure from \cite{boker07}.}
\end{center}
\end{figure}

\begin{figure}[t]
\begin{center}
\includegraphics[angle=0, width=.48\textwidth]{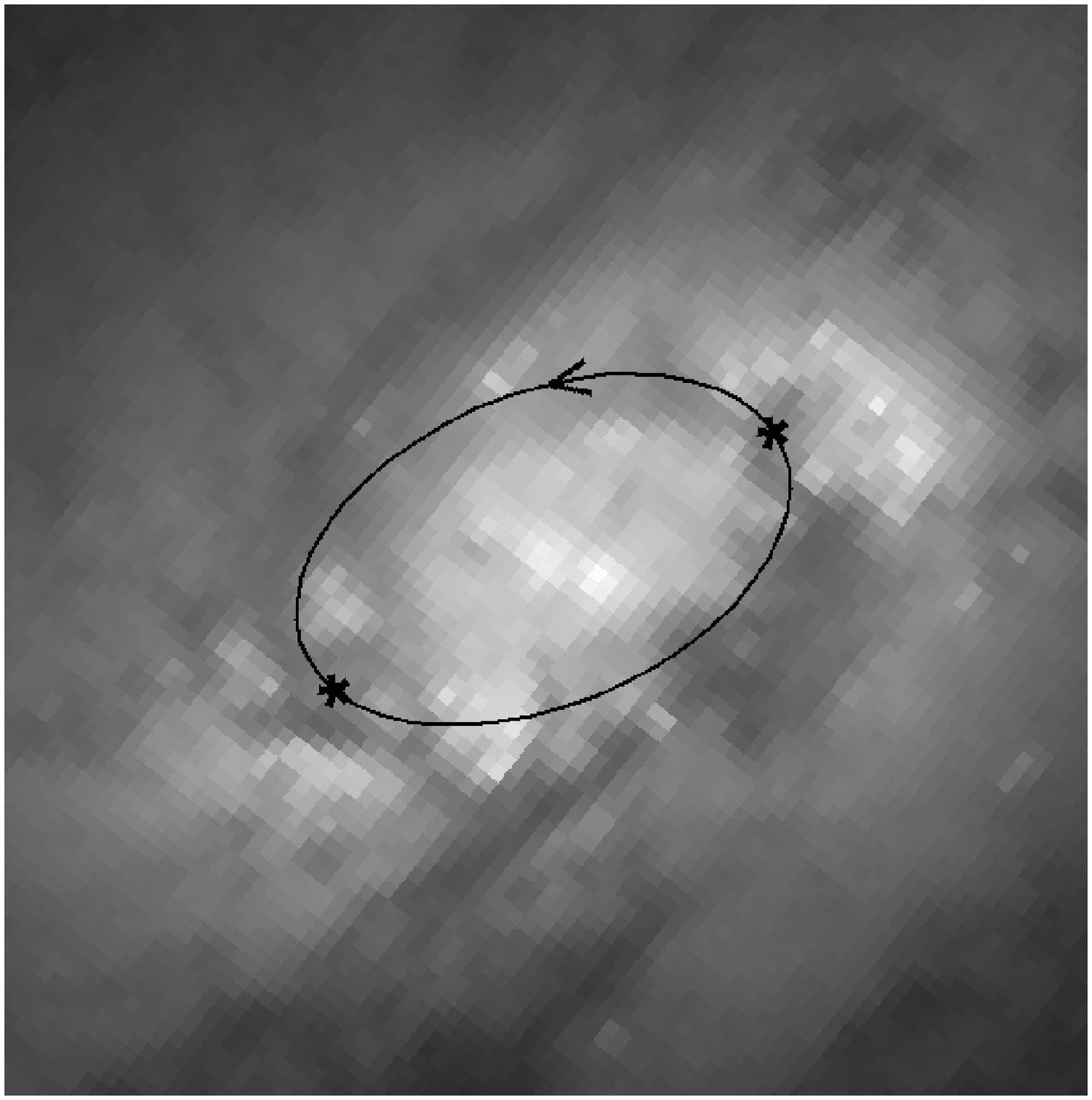}
\includegraphics[angle=0, width=.48\textwidth]{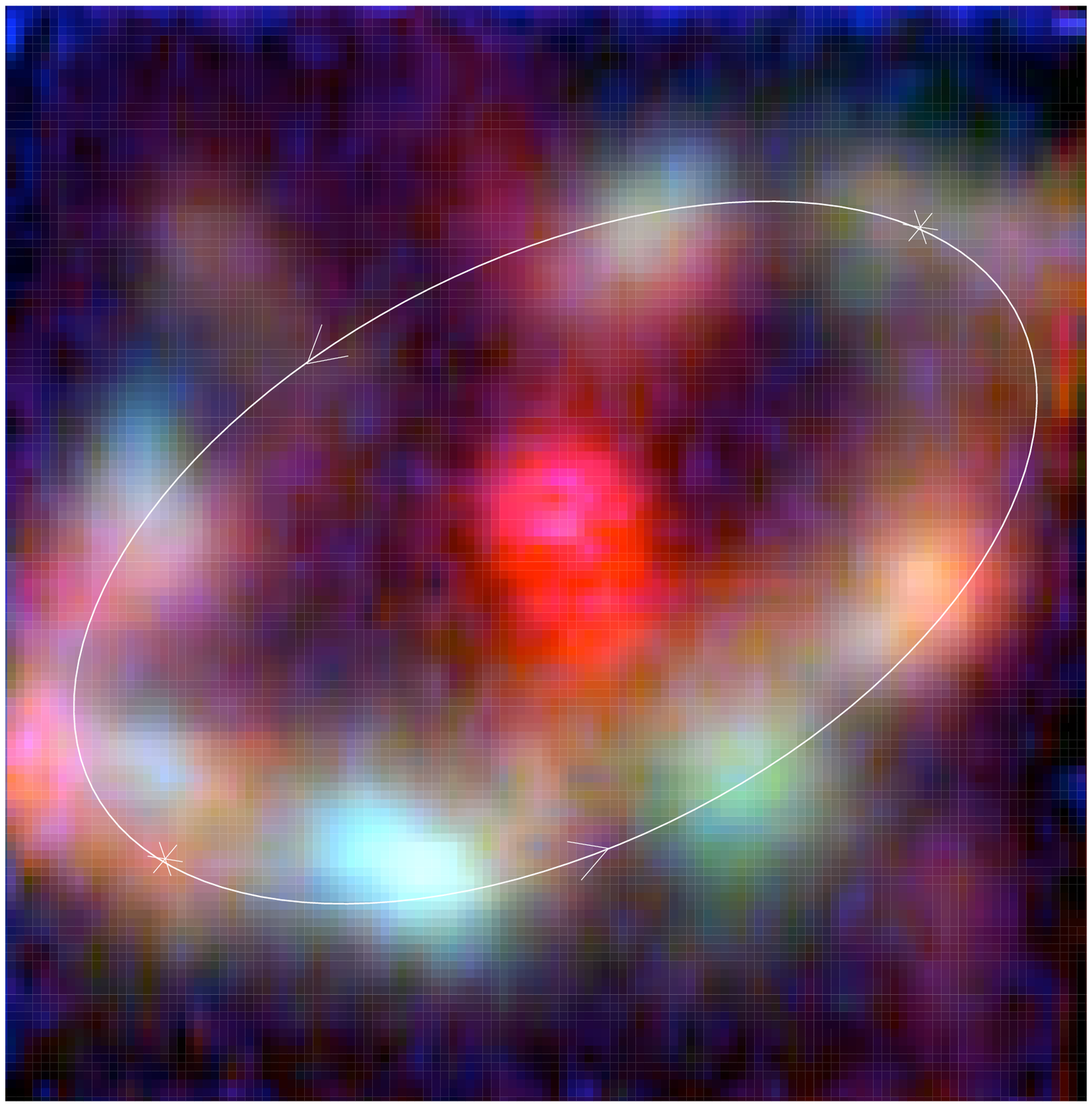}
\caption{Left: Archival HST/WFPC2 F606W map of NGC\,613. The outlined ellipse
marks the location and sense of rotation (denoted with and arrow) of matter
along the star-forming ring. Indicated are the assumed locations of the ODRs 
(marked with a star symbol)  where infalling gas is compressed. Right:
false color map of the emission line morphology in the inner 8''$\times$ 8'' of 
NGC\,613. The blue channel represents He{\sc I}, green Br$\gamma$, 
and red [Fe{\sc II}] emission. Note that - at least in the southern half of the ring 
- the hottest stars (as marked by He I emission in blue) are located immediately
downstream from the contact points. In contrast, the [Fe{\sc II}] emission in red
occurd predominantly further along the ring, with Br$\gamma$ emission 
dominated the region in between. This map supports the {\it 'Pearls on a string'} 
scenario, where star formation is triggered at the ODRs, forms stellar clusters 
than then age passively along their orbit. Figure from \cite{boker07}.}
\end{center}
\end{figure}

\section{The 'Popcorn' vs 'Pearls on a string' model scenarios}
In Figure~2 we illustrate two plausible mechanisms that describe how star-formation
proceeds along the cirnumnuclear rings. In the first scenario, the {\it 'Popcorn'}
model, gas enters the ring and accumulates around it with no preferred location. 
Once a critical density is reached, the gas becomes unstable to gravitational 
collapse and star formation is triggered. In this model, individual hot spots collapse 
at random times and locations within the ring, and therefore there is no 
systematic age sequence. In the figure (left plot) the different star bursts are 
denoted by the star symbols and the different colors indicate different star 
formation epochs. In the second scenario, Figure~2 (right), gas enters the ring 
at the intersection between the bar major axis and the inner Lindblad resonance. 
Downstream from this location, a point that we will call overdensity region (ODR), 
the gas density becomes sufficiently high to ignite star formation in a short-lived 
burst. A young cluster formed there will continue its orbit around the ring, but 
star formation will cease as soon as the first supernova explosions expel the 
gas. A series of starbursts triggered in the ODR will then produce a sequence 
of star clusters that enter the ring like {\it 'Pearls on a string'}. In this scenario, 
the star clusters should show a bi-polar age gradient around the ring, with the 
youngest clusters found close to the ODR, and increasingly older cluster ages 
in the direction of rotation, up to the opposite ODR. This is shown in the figure 
where the color sequence blue-green-red denotes clusters with increasing age.

In Figure~3 we put these two models to the test by showing the observations 
for NGC\,613, one of the five galaxies in our sample. In the left panel we display 
a F666W HST image of the nuclear regions. The ring is oulined with an ellipse, 
and two star symbols mark the position of the ODRs (based on the regions with 
highest dust extinction in the ring). In the right panel we present a false-color 
image constructed from the {\tt SINFONI} emission line maps of He{\sc I}, 
Br$\gamma$, and [Fe{\sc II}] assigned to the blue, green, and red color channels, 
respectively. The bottom half of the ring shows the trends expected under the 
{\it 'Pearls on a string'} picture: an age sequence (blue-green-red) in the 
different hotspots. The sequence is less obvious in the top part of the ring due 
to the strong interaction between the ring and the presence of a well-known radio 
jet \citep{hummel87}. In the whole sample, not shown here, three out of five 
galaxies show some evidence for an age gradient of hot spots along the ring, 
while the remaining two galaxies have incomplete information and thus are 
consistent with either model. 

At this point our results are not totally conclusive, although they favour the 
{\it 'Pearls on a string'} model. Larger samples are needed to shed 
more light on this issue. Nevertheless, as demonstrated in this contribution, the 
proposed method of using NIR line ratios to estimate relative ages of young star 
clusters is a very powerful tool for this purpose.\looseness-2
 
\begin{acknowledgments}
JFB would like to thank the organizing commetee and in particular Martin Bureau 
for a very fruitful and stimulating meeting.
\end{acknowledgments}



\begin{thebibliography}{}

\bibitem[Athanassoula (1992)]{atha92}
      {Athanassoula, E.} 1992, \textit{MNRAS}, 259, 345

\bibitem[B\"oker \etal\ (2007)]{boker07}
     {B\"oker, T., Falc\'on-Barroso, J., Schinnerer, E., Knapen, J.~H., Ryder, S.} 2007, \textit{AJ}, submitted

\bibitem[Combes \& Gerin (1985)]{cg85}
     {Combes, F., Gerin, M.} 1985, \textit{A\&A}, 150, 327

\bibitem[Hummel \etal\ (1987)]{hummel87}
     {Hummel, E., J\"ors\"ater, S., Lindblad, P. O., Sandqvist, A.} 1987, \textit{A\&A}, 172, 51

\bibitem[Knapen \etal\ (1995)]{knapen95}
     {Knapen, J. H., Beckman, J. E., Heller, C. H., Shlosman, I., de Jong, R. S.} 1995, \textit{ApJ}, 454, 623

\bibitem[Leitherer \etal\  (1999)]{sb99}
     {Leitherer, C., et al.} 1999, \textit{ApJS}, 123, 3

\bibitem[Schwarz (1984)]{schwarz84} 
     {Schwarz, M.~P.} 1984, \textit{MNRAS}, 209, 93 


\end{thebibliography}
\end{document}